\documentclass[aps,twocolumn,showpacs,preprintnumbers,prb,amsmath,amssymb,amsfonts,superscriptaddress,floatfix]{revtex4-1}

\usepackage[latin1]{inputenc}
\usepackage{graphics}
\usepackage{color}
\usepackage{amssymb}
\usepackage{amsmath}
\usepackage{overpic}
\usepackage{epstopdf}
\usepackage{bm}
\usepackage{graphicx}
\usepackage{subfigure}
\begin{document}

\title{Characterization of
rattling in relation to thermal conductivity:
ordered half-Heusler semiconductors}

\author{Zhenzhen Feng}
\affiliation{Key Laboratory of Materials Physics,
Institute of Solid State Physics, Chinese Academy of Sciences,
Hefei 230031, China}
\affiliation{Department of Physics and Astronomy, University of Missouri,
Columbia, MO 65211-7010, USA}
\affiliation{Institute for Computational Materials Science,
School of Physics and Electronics, Henan University, Kaifeng, 475004, China}

\author{Yuhao Fu}
\affiliation{Department of Physics and Astronomy, University of Missouri,
Columbia, MO 65211-7010, USA}
\affiliation{College of Physics,
Jilin University, Changchun 130012, China}

\author{Yongsheng Zhang}
\email{yshzhang@theory.issp.ac.cn}
\affiliation{Key Laboratory of Materials Physics,
Institute of Solid State Physics, Chinese Academy of Sciences,
Hefei 230031, China}
\affiliation{Science Island Branch of Graduate School,
University of Science and Technology of China, Hefei 230026, China}

\author{David J. Singh}
\email{singhdj@missouri.edu}
\affiliation{Department of Physics and Astronomy, University of Missouri,
 Columbia, MO 65211-7010, USA}
\affiliation{Department of Chemistry, University of Missouri,
 Columbia, MO 65211, USA}

\date{\today}

\begin{abstract}
The factors that affect the thermal conductivity of semiconductors
is a topic of great scientific interest,
especially in relation to thermoelectrics.
Key developments have been the concept of the phonon-glass-electron-crystal
(PGEC) and the related idea of rattling to achieve this.
We use first principles phonon and thermal conductivity calculations
in order to explore the concept of rattling
for stoichiometric ordered half-Heusler compounds.
These compounds can be regarded as filled zinc blende materials,
and the filling atom could be viewed as a rattler if it is weakly bound.
We use two simple metrics, one related to the frequency and the other
to bond frustration and anharmonicity. We find that both measures correlate
with thermal conductivity.
This suggests that both may be useful in screening materials
for low thermal conductivity.
\end{abstract}

\maketitle

\section{Introduction}

Thermal conductivity is an important primary quantity in describing the
behavior of a material.
It is of particular importance for thermoelectrics (TE),
where low thermal conductivity, especially
low lattice thermal conductivity is desired.
The conversion efficiency is characterized by the thermoelectric
figure of merit $ZT=\sigma S^{2}T/\left(\kappa_{e}+\kappa_{l}\right)$,
where $S$ is the Seebeck coefficient, $\sigma$ is the electrical conductivity,
$\kappa_{e}$ is the electronic thermal conductivity,
$\kappa_{l}$ is the lattice thermal conductivity,
and $T$ is the absolute temperature.
\cite{wood,he2017advances}
The search for high $ZT$
led to the concept of the phonon-glass-electron crystal (PGEC),
which is the idea of looking for semiconductors that have low
electron scattering, and therefore high electrical conductivity, but at
the same time very strong phonon scattering.
\cite{tritt1999holey,beekman}
This together with the idea of rattling has been an influential
and successful theme in TE research. It has led to the
identification of many interesting novel materials, including
clathrates and filled skutterudites.
\cite{tritt1999holey,cohn,takabatake2014phonon,sales,nolas1996effect,shi}

The purpose of this paper is to examine this concept in ordered
stoichiometric half-Heusler compounds.
The half-Heusler structure can be viewed as a filled zinc blende
lattice.
Therefore, if one of the atoms is weakly bound, one could imagine that
it may serve as a rattler lowering the thermal conductivity.
However, as discussed below, application of this concept to half-Heuslers
is non-trivial, since they have rather complex lattice dynamics.
Nonetheless, we do find that some compounds can be described as having
rattling behavior in relation to thermal conductivity,
and we discuss its characterization using two different measures.

The basic idea of rattling is to start with a semiconductor framework
and fill with guest atoms that might strongly scatter
heat carrying phonons of the host lattice, while maintaining
the electronic structure and electronic transport.
Realizations have invariably involved guest atoms
that are bound in the host by chemical interactions, for example,
bonding of the fillers in skutterudites,
\cite{nolas1998,luo}
leading to modifications of the electronic structure.
However, with careful selection
these electronic changes can be beneficial for thermoelectric performance
beyond the reduced $\kappa_l$.
\cite{singh1997,shi2015,yang-review}

This raises the questions of mechanism, how
to identify rattling in a material, and how effective a given
rattler may be in reducing thermal conductivity.
Lattice thermal conductivity in normal crystalline materials
is governed by a dispersion relation, i.e. the phonons,
and scattering, which can have different contributions.
For clean
materials anharmonic three phonon umklapp scattering is often dominant
in controlling thermal conductivity.
\cite{lindsay}

Guest atoms serving as rattlers
may introduce low frequency vibrations that hybridize with
the heat carrying acoustic modes. This will reduce the group velocity
in the frequency range near the crossings of the acoustic branches with the
guest atom vibrations due to hybridization with the rattler optic phonons.
The resulting reduction in thermal conductivity is then a harmonic effect,
arising from changes in dispersion due to hybridization of the vibrations
of the host lattice and the guest atoms.
In this view, the frequency of the rattling vibrations and the
harmonic interaction with the host lattice plays the central role. This
view has led to the development of some of the high performance
skutterudites, where multiple fillers scatter acoustic phonons in
different frequency ranges.
\cite{shi}

Another view is that strong anharmonicity associated with weak bonding
is crucial. The strong anharmonicity then leads strong phonon
scattering. Rattling systems often have weakly bonded atoms with
long, stretched bonds and large atomic displacement parameters
(large mean square displacements), which
have been used to characterize this type of behavior.
\cite{sales-1999}

These two views (harmonic interactions mixing modes vs.
strong anharmonicity due from weak bonding) are seemingly divergent,
and have been the source of some controversy, e.g. in skutterudites.
\cite{keppens,feldman2000}
They are in fact different.
However, they are complementary in that both mechanisms may be
operative in a given system, and both may be ways of achieving low
thermal conductivity.

Importantly, recently developed thermal conductivity methods based
on anharmonic phonon scattering allow one to directly calculate
lattice thermal conductivity. 
\cite{lindsay,li2012thermal,li2014shengbte,togo,hellman}
Here we use these tools in conjunction with simple parameters that can
be extracted from phonon dispersions to explore these different views and
to find phonon based metrics that may be useful in identifying rattling and
low thermal conductivity without direct thermal conductivity calculations.

As mentioned, we use half-Heusler semiconductors for this purpose.
This is a very large class of compounds that has a simple crystal structure
and contains many known good thermoelectric materials.
\cite{tritt1999holey,uher1999transport,joshi2011enhancement,lee2011electronic,gurth,zhou,samsonidze,zhu,zhu2}
This family shows considerable chemical flexibility,
as reflected in the large number of compounds.
Half-Heusler compounds also exhibit a very large
range of thermal conductivities.
\cite{RN2359}
The structure can be regarded as a filled
zinc blende structure, which suggests possibilities for rattling
if the filling atom is weakly bound.
The interest in half-Heusler TE materials has motivated much
work on and current interest in
their thermal conductivities and ways of minimizing them.
\cite{RN2359,shiomi,culp2006effect,chen2013effect,ding,holuj,eliassen,gong,berland}
Furthermore, as recently discussed by Berland and co-workers,
\cite{berland} their thermal
properties are very subtle. Besides the large range of thermal
conductivities that can occur, the low thermal conductivity compounds
of interest for thermoelectrics,
can have sizable
reductions in lattice thermal conductivity due to disorder and grain
boundaries, as well as 
substantial electronic contributions,
\cite{berland}
which are, however, highly non-trivial to extract from experimental data
alone. \cite{putatunda}
In addition, there are several mechanisms that can be important
for reducing the thermal conductivity of half-Heusler compounds.
These include site disorder, alloy scattering, and anharmonicity related
to lone pair physics (also discussed as resonant bonding),
\cite{lee}
and other features of the bonding that lead to anharmonicity.
\cite{berland,hermet}
Here we examine the concept of rattling in order to characterize it in the
context of these materials and to examine the extent to which and
how this concept can be applied.

\section{Structure and Methods} 

\begin{figure}
\includegraphics[width=\columnwidth]{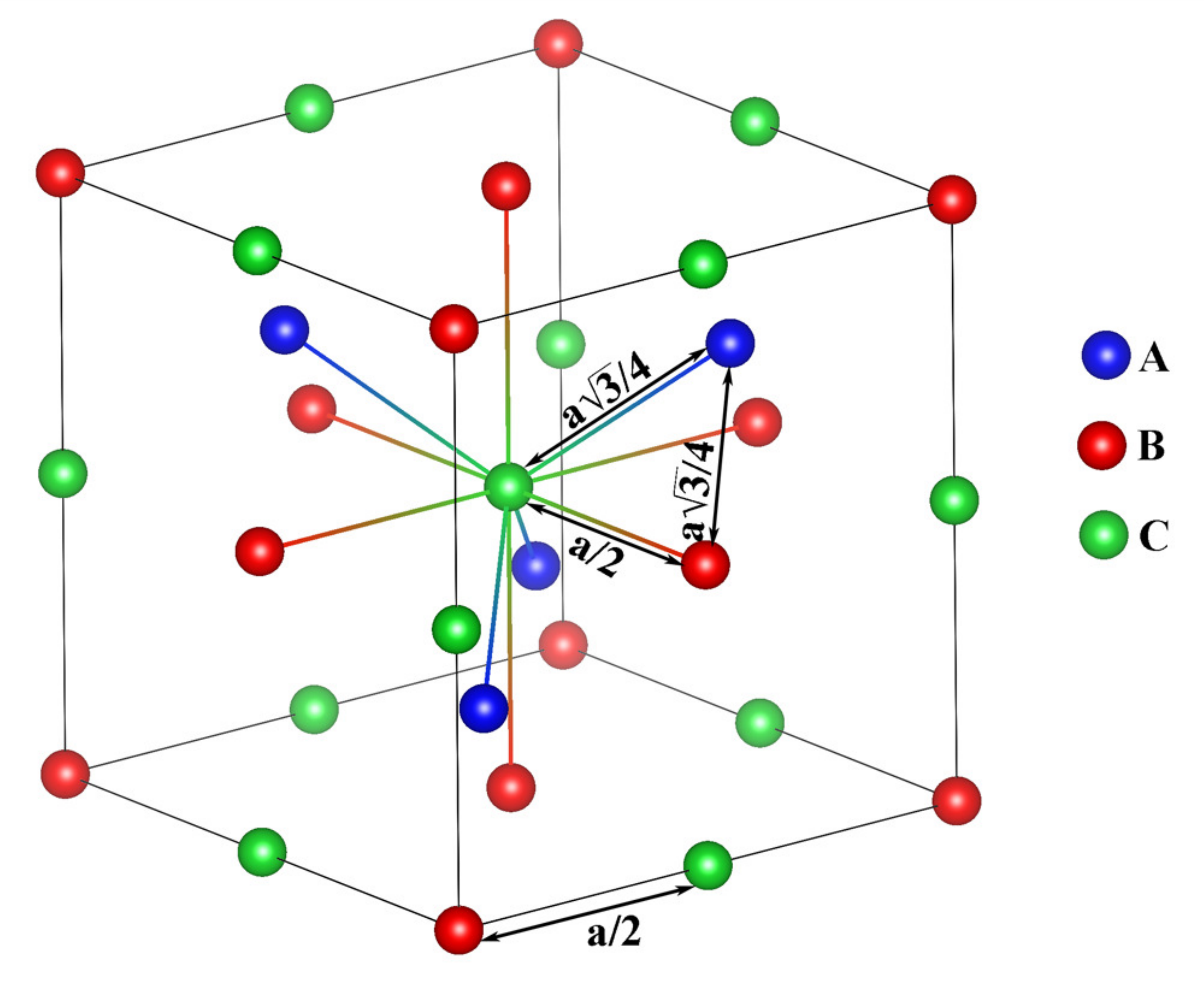}
\caption{\label{figure1} The half-Heusler crystal structure,
showing the bond lengths.
Note in particular the two bond identical bond lengths of $\sqrt{3}a/4$
even though the atoms involved are different.}
\end{figure}

Half-Heuslers are ternary intermetallics with general formula ABC,
occurring in cubic space group
$\mathrm{F} \overline{4} 3 \mathrm{m}$
(Figure \ref{figure1}).
The structure consists of three interpenetrating face centered cubic (fcc)
sublattices and one vacant fcc sublattice, which if filled would yield
the full-Heusler structure.
Here we follow the common notation, where the
A occupies 4c (1/4, 1/4, 1/4),
B occupies 4a (0, 0, 0)
and C occupies 4b (1/2, 1/2, 1/2), rather than IUPAC notation in order
to clearly connect with the structures.
In terms of coordination, each A atom has four B neighbors and four C
neighbors at distance $a \sqrt{3}$/4.
The B and C atoms are each coordinated by four A neighbors at a distance of
$a\sqrt{3}$/4.
We note that exchanging the A atom with B or C results in a different
material.
Here, we checked each possibility using the total
energy and performed calculations for the lowest energy ordering.
The results are consistent with the report of Carrete and coworkers.
\cite{RN2359}

The half-Heusler structure has a single structural parameter,
the lattice parameter
$a$, but two distinct short nearest neighbor bonds. This plus
the wide chemical flexibility of the structure type provides
opportunities for having compounds with frustrated bond lengths, and therefore
potentially rattling atom physics.
However, simple measures of bond satisfaction,
e.g. comparisons of sums of ionic radii with bond lengths,
as have been highly successful in understanding the structures of oxides,
are difficult to apply in
half-Heusler compounds because of the variety of different bonding
types (metallic, covalent, ionic and mixtures) that occur in this family.
\cite{graf,bende,RN2430}
Goals of the present work include finding ways of describing rattling
based on the idea of bond length frustration and examining the
extent to which these measures, and therefore the idea of rattling
due to bond frustration underlies the exceptional range of thermal
conductivities found in half-Heusler semiconductors.

For this purpose we use a set of 75 known and potential half-Heusler
compounds that were identified by Carrete and co-workers,
\cite{RN2359}
and were previously used in other screens related to thermoelectrics, including
thermal and electronic properties, and searches for other systems.
\cite{feng,legrain}
This list includes both known and hypothetical materials predicted by
stability analysis.

We calculated the phonon dispersions for all 75 of the proposed
half-Heusler compounds from the data set.
We find that 74 of them have phonon dispersions with only stable modes, and
use these compounds as the set to analyze.
SbNaSr is found to be dynamically unstable.
We then directly calculated the lattice thermal conductivity of
each compound by solving
the linearized Boltzmann-Peierls transport equation
with the ShengBTE package.
This gives us a set of phonon calculations, which are the basis for
our analysis, and a set of thermal conductivities. For some compounds
that had low thermal conductivities, we did additional calculations
using the temperature dependent effective potential method to generate the
anharmonic coefficients for the Boltzmann transport calculations. This
provides a more stable and presumably more accurate result
for such cases, but uses the same underlying three-phonon scattering
physics for the thermal conductivity.

Our density functional calculations were done using the projector
augmented wave (PAW) method,
\cite{RN245}
as implemented in the VASP code.
\cite{RN2407}
We used the generalized gradient approximation
of Perdew, Burke and Ernzerhof (PBE-GGA), with an energy
cutoff of 500 eV and Brillouin zone samples based on
a 10$\times$10$\times$10 mesh.

The thermal conductivity calculations were done with standard iterative
solution of the Boltzmann equation
\cite{omini1996beyond} with harmonic and anharmonic
interatomic force constants from density functional theory.
All contributions from two-phonon and three-phonon
scattering processes were included.
For the cubic systems considered here,
$\kappa_{l}$ is a scalar quantity given by
 \begin{equation}
\kappa_{l} \equiv \kappa_{l}^{\alpha \alpha}=\frac{1}{N V} \sum_{\lambda} C_{\lambda} v_{\lambda}^{\alpha} v_{\lambda}^{\alpha} \tau_{\lambda}
\end{equation}
Where $\lambda$ denotes a phonon mode in branch $\textit{p}$ with wave vector $\textbf{\textit{q}}$, 
$N$ is the number of uniformly spaced $\textbf{\textit{q}}$ points
in the phonon BZ, $V$ is the volume of the unit cell,
$C_{\lambda}$ is the specific heat,
$v_{\lambda}$ is the phonon group velocity, and 
$\tau_{\lambda}$ is the lifetime
with an applied temperature gradient along the $\alpha$ direction. 
$\tau_{\lambda}$ is determined by the processes of two-phonon scattering
from isotopic disorder and three-phonon anharmonic scattering.
Here we focus on three-phonon anharmonic scattering.
$\tau_{\lambda}$ is given by the sum of all possible transition
probabilities for mode $\lambda$ with modes
$\lambda^{\prime}$ and $\lambda^{\prime \prime}$,

\begin{equation}
\begin{aligned}
\Gamma_{\lambda \lambda^{\prime} \lambda^{\prime \prime}}^{ \pm}= & \frac{\hbar}{8 N_{0}}\left\{\begin{array}{c}{n_{\lambda^{\prime}}^{0}-n_{\lambda^{\prime \prime}}^{0}} \\ {n_{\lambda^{\prime}}^{0}+n_{\lambda^{\prime \prime}}^{0}+1}\end{array}\right\} \\
& \times \left|\Phi_{\lambda \lambda^{\prime} \lambda^{\prime \prime}}\right|^{2} \frac{\delta\left(\omega_{\lambda} \pm \omega_{\lambda^{\prime}}-\omega_{\lambda^{\prime \prime}}\right)}{\omega_{\lambda} \omega_{\lambda^{\prime}} \omega_{\lambda^{\prime \prime}}} \end{aligned}
\end{equation}

\noindent 
that satisfy momentum and energy conversation. where the upper (lower) row
is curly brackets go with the +(-) signs are
for absorption (emission) processes.
$\omega_{\lambda}$ is the angular frequency
corresponding to the $\lambda$th mode,
and $\left|\Phi_{\lambda \lambda^{\prime} \lambda^{\prime \prime}}\right|^{2}
$ are the scattering matrix elements,

\begin{equation}
\begin{aligned} \Phi_{\lambda \lambda^{\prime} \lambda^{\prime \prime}}=
& \sum_{k} \sum_{l^{\prime} k^{\prime}} \sum_{l^{\prime \prime} k^{\prime \prime}} \sum_{\alpha \beta \gamma} \Phi_{\alpha \beta \gamma}\left(0 k, l^{\prime} k^{\prime}, l^{\prime \prime} k^{\prime \prime}\right) \\
& \times \frac{e_{\alpha k}^{\lambda} e_{\beta k^{\prime}}^{\lambda^{\prime}} e_{\gamma k^{\prime \prime}}^{\prime \prime}}{\sqrt{M_{k} M_{k^{\prime}} M_{k^{\prime \prime}}}} e^{i q^{\prime} R_{l^{\prime}}} e^{i q^{\prime \prime} R_{l^{\prime \prime}}} \end{aligned}
\end{equation}

\noindent
where $M_{k}$ is the atomic mass of the $k$th atom,
and $\Phi_{\alpha \beta \gamma}\left(0 k, l^{\prime} k^{\prime},
l^{\prime \prime} k^{\prime \prime}\right)$ are the
anharmonic interatomic force constants (IFCs).
Then the phonon angular frequencies $\omega_{\lambda}$ are
obtained from diagonalization of the dynamical matrix.

\begin{table*}[tbp]
\caption{Calculated lattice thermal conductivities of
74 half-Heusler compounds at 300 K, in units of W/mK}
\label{thermal}\tabcolsep0.01in
\begin{ruledtabular}
\renewcommand{\arraystretch}{1.1}
\begin{tabular}{cccccccccccccc}
  
Comp.   & $\kappa_{l}$  &Comp. & $\kappa_{l}$  &Comp. & $\kappa_{l}$ &Comp. & $\kappa_{l}$  \\ \hline
\hline
PtLaSb  & 0.84   & PdHfSn & 14.76 & RuTeZr  & 19.61 & CoGeNb  & 26.99  \\
RhLaTe  & 1.21   & NiAsSc & 15.76 & CoNbSn  & 20.06 & OsSbTa  & 28.12   \\
BiBaK   & 2.26   & NiPbZr & 15.94 & CoSbTi  & 20.34 & IrGeV   & 28.35   \\
PCdNa   & 2.27   & RhAsZr & 16.22 & RuSbTa  & 20.45 & IrGeNb  & 28.71   \\
ZnLiSb  & 7.03   & CoBiZr & 16.70 & CoGeTa  & 20.73 & CoSiTa  & 28.87    \\
CoAsTi  & 7.28   & CoSnV  & 16.83 & CoAsZr  & 20.82 & OsNbSb  & 29.38    \\
NiBiY   & 7.44   & RhSnTa & 16.99 & NiGeTi  & 20.88 & FeNbSb  & 29.61   \\
IrAsZr  & 7.70   & CoNbSi & 17.06 & PtGeTi  & 21.61 & FeSbTa  & 30.28   \\
PdBiSc  & 7.82   & PdGeZr & 17.13 & IrAsTi  & 21.79 & IrGeTa  & 32.19   \\
PdPbZr  & 9.42   & FeGeW  & 17.16 & CoSnTa  & 21.79 & AuAlHf  & 33.85   \\
CoHfSb  & 10.01  & SiAlLi & 17.19 & NiGeZr  & 22.28 & RuAsNb  & 36.08    \\
RhBiHf  & 10.46  & RhNbSn & 17.57 & NiGaNb  & 22.84 & FeAsNb  & 37.58   \\
CoSbZr  & 10.96  & GeAlLi & 17.71 & FeSbV   & 23.01 & BLiSi   & 37.71   \\
IrBiZr  & 11.15  & CoAsHf & 17.95 & CoGeV   & 24.30 & IrSnTa  & 78.09    \\
RhBiTi  & 11.41  & NiGeHf & 18.08 & FeAsTa  & 24.55 & &  \\
NiBiSc  & 11.56  & NiHfSn & 18.28 & PtGaTa  & 24.55 & &  \\
RhBiZr  & 12.45  & IrNbSn & 18.62 & RuNbSb  & 24.91 & &  \\
PtGeZr  & 12.68  & CoBiHf & 18.74 & RhAsTi  & 24.94 & &  \\
NiSnZr  & 13.26  & IrHfSb & 18.96 & RuAsTa  & 25.58 & &  \\
NiSnTi  & 14.63  & FeTeTi & 19.47 & CoBiTi  & 26.67 & &  \\
\end{tabular}
\end{ruledtabular}
\end{table*}

The phonon dispersions and the harmonic second-order interatomic force
constants (IFCs) were calculated using the frozen phonon method,
as implemented in the Phonopy package. \cite{RN2408}
4$\times$4$\times$4 supercells (with 192 atoms in total)
and 2$\times$2$\times$2 supercell $k$-meshes were used for the
dynamic matrix.
As mentioned, SbNaSr is found
to be dynamically unstable, and is not further considered here.
The anharmonic IFCs were calculated using the same supercell and $k$-mesh.
The ShengBTE package \cite{li2014shengbte, li2012thermal}
was employed to iteratively solve the phonon Boltzmann equation.
15$\times$15$\times$15
{\bf{q}}-grids were used. 
In additional for low thermal conductivity compounds, which are
of particular interest for this study, we did
ab initio molecular dynamics (AIMD).
This allows determination of temperature dependent 
harmonic and anharmonic interatomic force constants.

This can be different from static calculations for highly anharmonic materials,
i.e. materials that have low thermal conductivity due to strong anharmonicity.
We calculated thermal conductivities for
PtLaSb, SiAlLi, BiBaK, PCdNa, CoAsHf, PdBiSc, GeAlLi, and ZnLiSb
using this temperature-dependent effective potential (TDEP) method.
\cite{hellman2013temperature}
For this purpose we employed
Born-Oppenheimer molecular dynamics with the
PAW method, as implemented in the VASP code. The parameters were similar
to the ground state calculations, except that a somewhat lower planewave
cutoff of 330 eV was used.
The simulations were run for approximately 100 ps with a time step of 1 fs and
a temperature of 300 K, which is the temperature that we focus on in this
study.

The use of TDEP gives significant differences from
ShengBTE results for only
three compounds, specifically, SiAlLi, CoAsHf, and ZnLiSb.
Therefore, in the following we report the data calculated by
ShengBTE except for these three compounds (SiAlLi, CoAsHf, and ZnLiSb),
for which TDEP results are used.
 
\section{Results and Discussion}

\begin{figure}
\includegraphics[width=0.85\columnwidth]{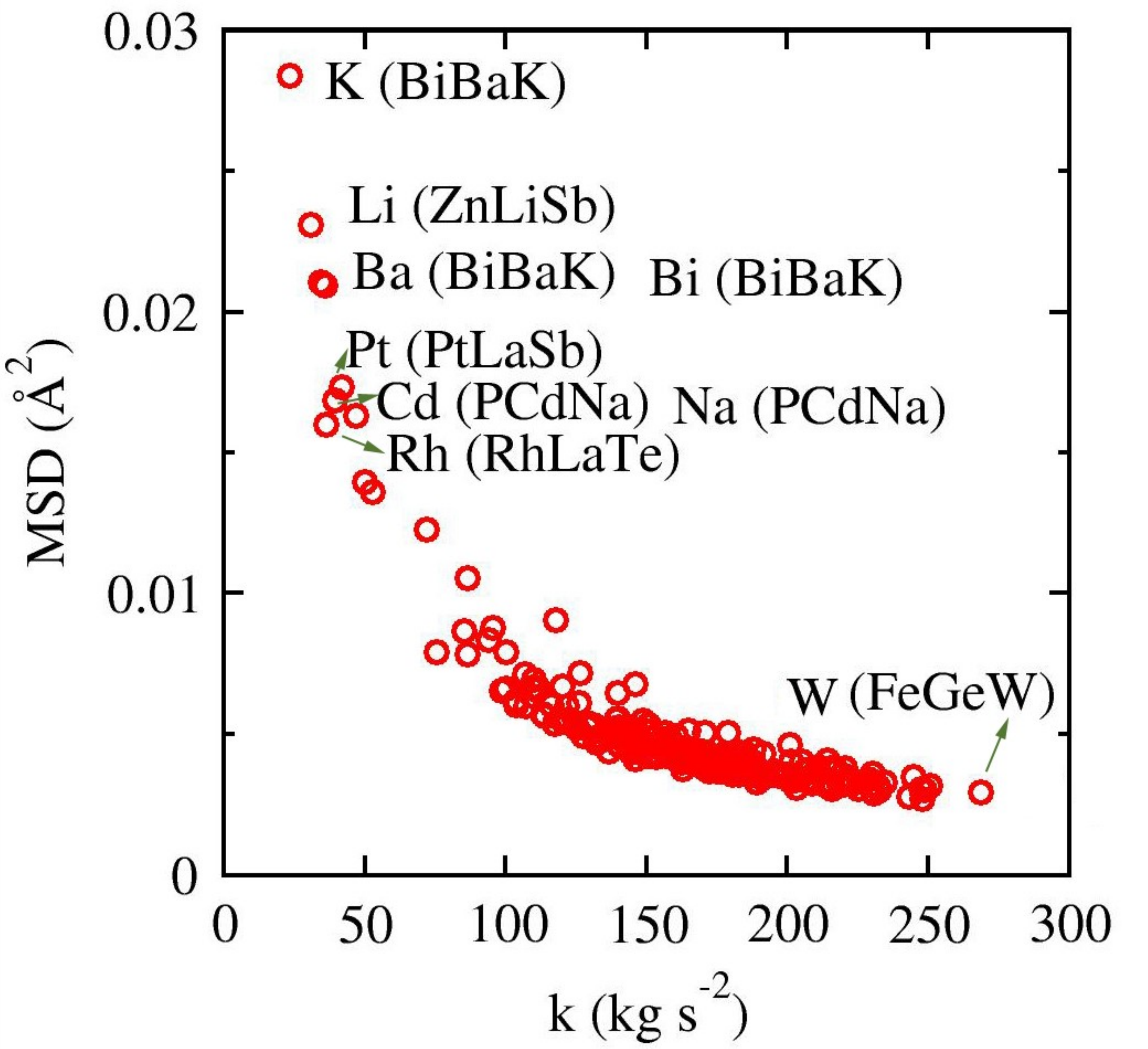}
\caption{\label{msd} Calculated mean square displacement (MSD)
at 300 K vs. effective spring constant for the atoms in our dataset. Labels denote specific atoms in
	the compounds in parentheses, e.g. Li (ZnLiSb) denotes the Li in ZnLiSb.}
\end{figure}

\begin{figure}
\includegraphics[width=\columnwidth]{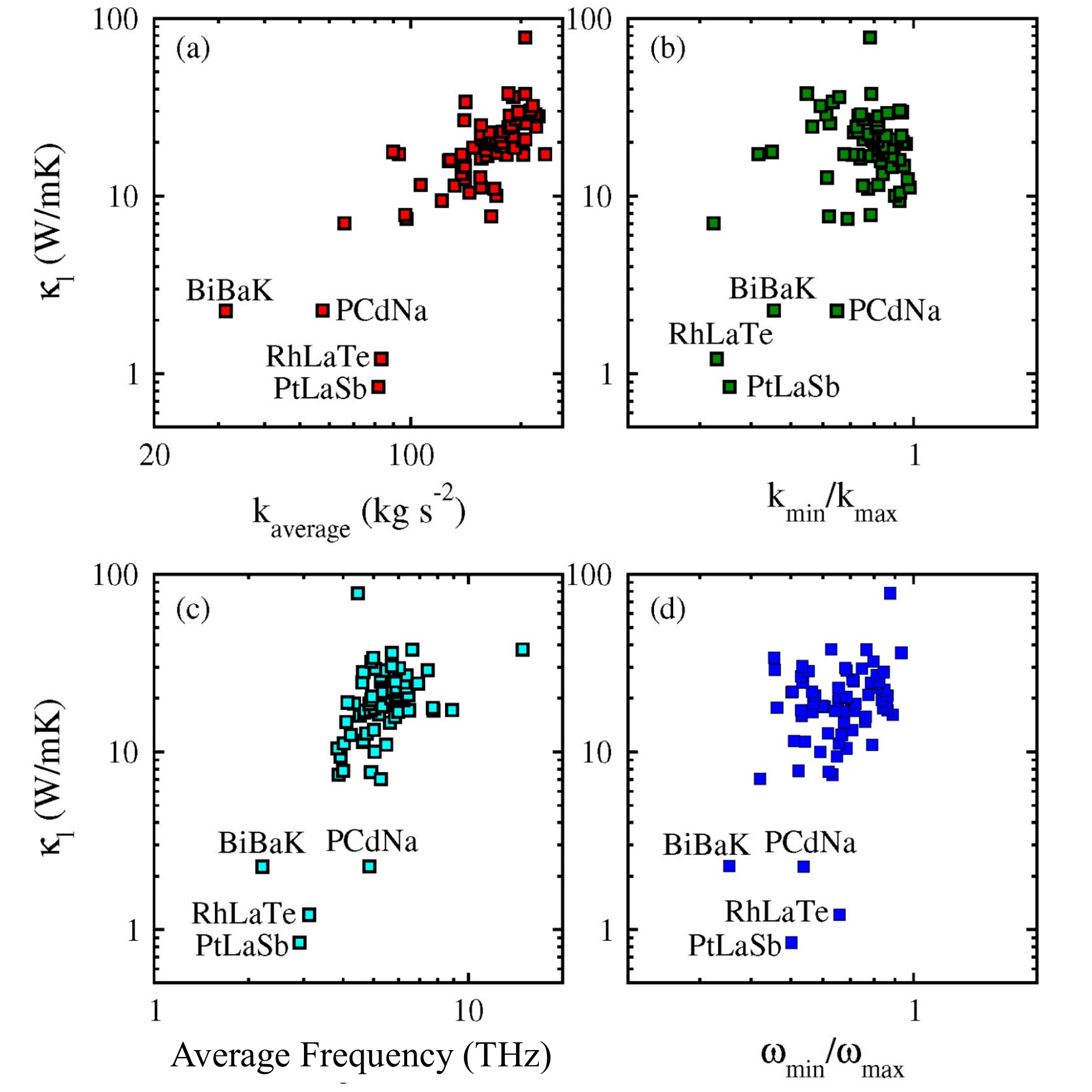}
\caption{\label{figure2} 
(a) The average spring constant,
$k_{average}$ as a function of lattice thermal conductivity
for half-Heuslers.
(b) The ratio of $k_{min}$ to $k_{max}$,
here $k_{min}$ and $k_{max}$ are the smallest
and largest spring constants among three atoms,
respectively.
(c) The average frequency,
and corresponding the lattice thermal conductivity
(here the average is calculated using the total DOS).
(d) The ratio of $\omega _{min}$ to $\omega _{max}$,
here $\omega _{min}$ and $\omega _{max}$ are the smallest
and largest angular frequency among three atoms.}
\end{figure}

The calculated thermal conductivities for the 74 half-Heusler compounds
are listed in Table \ref{thermal}.
This is the basis for the comparisons given in the remainder of this paper.
As seen in Table \ref{thermal}, four compounds have
very low 300 K lattice thermal conductivity,
PtLaSb, RhLaTe, BiBaK, and PCdN, with $\kappa_l$ of
0.84 W/mK, 1.21 W/mK, 2.26 W/mK, and 2.27 W/mK, respectively.

There are also many compounds with much higher thermal conductivity,
ranging up to 78 W/mK in the case of IrSnTa.
The wide range of thermal conductivities implies that thermal conductivity
of half-Heusler compounds involves rather rich physics related to
phonon scattering.
This wide range cannot be understood just by invoking measures based
on the masses of the atoms involved or the acoustic phonon group velocities.
For example, the lowest
thermal conductivity material,
PtLaSb, with $\kappa_l$ of 0.84 W/mK has a mass per unit cell of 
$M_{cell}$=455.75 amu,
while IrSnTa has $M_{cell}$=491.88 amu, i.e.
close to, and even slightly larger than that of PbLaSb.
Other measures are discussed below.

\begin{figure}
\includegraphics[width=0.7\columnwidth]{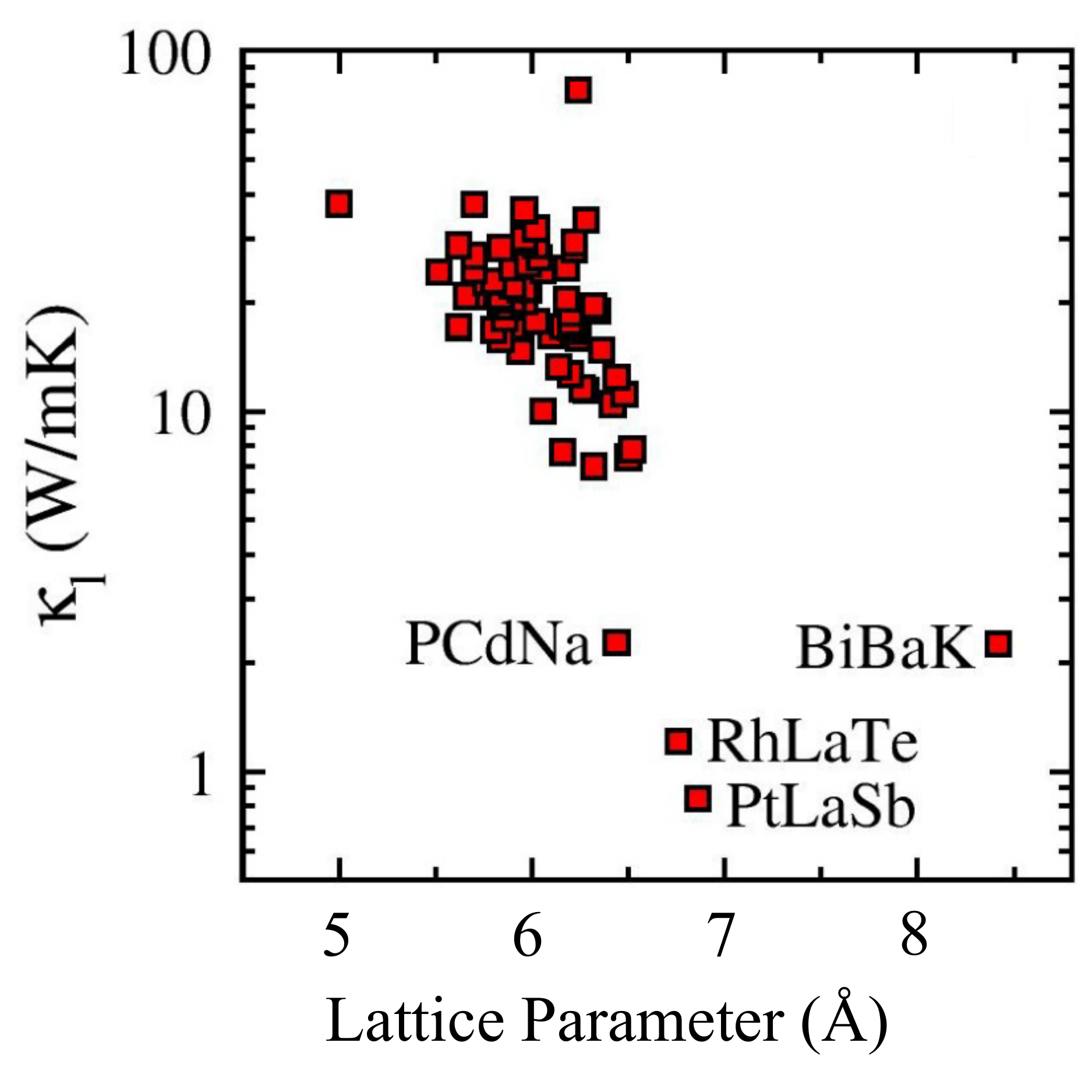}
\caption{\label{lattice}
Thermal conductivity and lattice parameter for half-Heusler compounds.}
\end{figure}

\begin{figure}
\includegraphics[width=0.7\columnwidth]{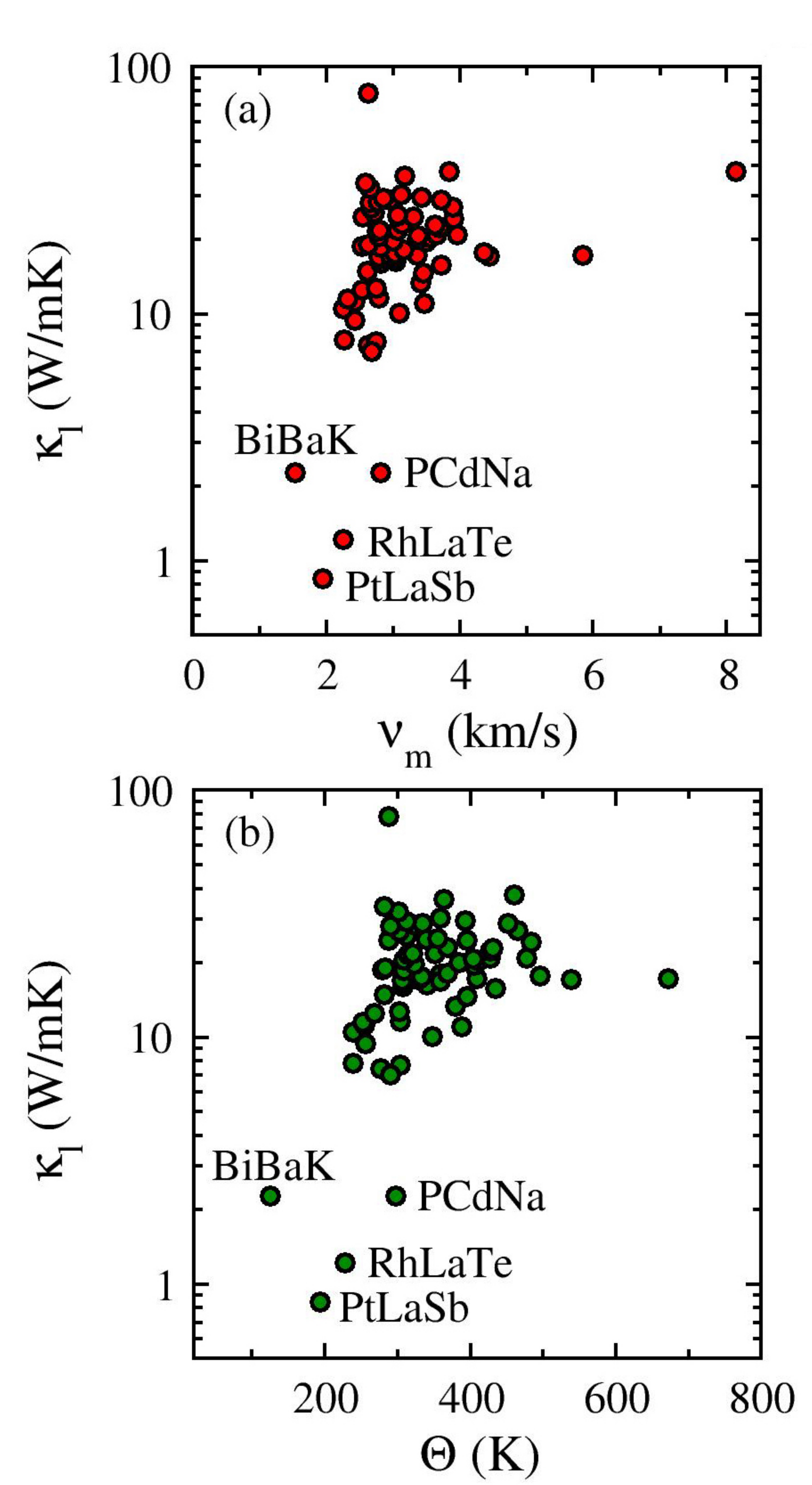}
\caption{\label{figure3}
Relationship of thermal conductivity with
(a) the calculated average sound velocities and
(b) the Debye temperature.}
\end{figure}

The half-Heusler structure contains three different atoms.
If one of the atoms is very small, bound loosely and/or provides
low optic phonon frequencies, it might be considered as a rattler.
We constructed two measures based on the phonon dispersions. We note
that phonon dispersions are much easier to calculate than thermal 
conductivities.
In the following we explore ways of using these to understand thermal
conductivity and identify materials with potentially low thermal conductivity.
These measures are based on the atom projected phonon density of states.
The first measure is from an average phonon frequency for each atom, calculated
using the first frequency moment of the projected density of states.
The idea is that if one atom has a much lower average frequency than the
other two, or than the average frequency of the solid, then it might be
a rattler.
This measure connects with the idea that associates rattling with
the introduction of low frequency optic branches than intersect with 
the heat carrying acoustic branches, scattering acoustic phonons and
modifying the harmonic
acoustic phonon dispersions to reduce thermal conductivity.
This point of view was emphasized in the context of skutterudites
by Feldman and others. \cite{feldman2000,shi}

A complementary point of view, discussed in the same context by
Keppens and co-workers, \cite{keppens}
is that strong anharmonicity is key.
This is more difficult to characterize from the harmonic phonon frequencies.
Here, we examine this using the concept where stretched bonds lead to
strong anharmonicity, and characterize stretched bonds by low force constants.
In particular, we compute an average spring constant for each atom,
$\alpha$, using
the average angular frequency, $\overline{\omega}_\alpha$,
obtained as above, so that $k_\alpha=\overline{\omega}^2m_\alpha$,
where $m_\alpha$ is the mass of the atom.
We also define an average spring constant, which is just
$k_{average}=(k_1+k_2+k_3)/3$. This is different from the average that
would be obtained from the average phonon frequency, due to the
power of two in the formula for the spring constant.

As mentioned, large mean square displacements have been associated with
rattling. All sites in the half-Heusler structures have cubic site
symmetry. Therefore at the harmonic level the mean square displacements
are isotropic.
The mean square displacements (MSD) are expected to be closely
related to the effective spring constant, MSD$\propto$ 1/$k_\alpha$
due to equipartition at temperatures high enough
that quantum effects are not important.
This in fact is the case, as shown in Fig. \ref{msd}, where the
MSD for all atoms in the data set is plotted vs. effective spring
constant from the phonon density of states.

Fig. \ref{figure2}(a) shows the thermal conductivity versus the $k_{average}$.
As may be expected,
$k_{average}$ has a positive correlation with $\kappa_{l}$, which
is simply a reflection of the fact that a stiff lattice favors high 
thermal conductivity.
This is simply understood in terms of the Callaway model, \cite{callaway}
where thermal
conductivity is proportional to a product of specific heat,
phonon group velocity and phonon mean free path.
A stiff lattice yields high phonon frequencies, and therefore high phonon group
velocities. This leads to high
$\kappa_l$ due to the proportionality of $\kappa_l$ and velocity.

A similar correlation is seen with the average frequency, though the separation
of the high thermal conductivity compounds is actually somewhat weaker, even
though it would seem at first glance that it should be better due to the
more direct connection of frequency with the Callaway picture.
The difference between these is in the mass, in other words the fact that
lattices with heavy atoms have lower phonon frequencies and sound velocities,
for the same force constants.
This suggests that in fact force constants are important, perhaps because
they also may contain chemical information about anharmonicity.
As mentioned above, the mass itself does not explain the wide range 
of thermal conductivities.
We also note that there is a correlation between thermal conductivity and the
lattice parameter, as shown in Fig. \ref{lattice}. This
reflects the idea that larger lattice parameter corresponds to weaker
bonding in general.
None of these correlations is strong enough to be used as a
reliable predictor of thermal conductivity by itself.

\begin{figure*}
\includegraphics[width=0.85\textwidth]{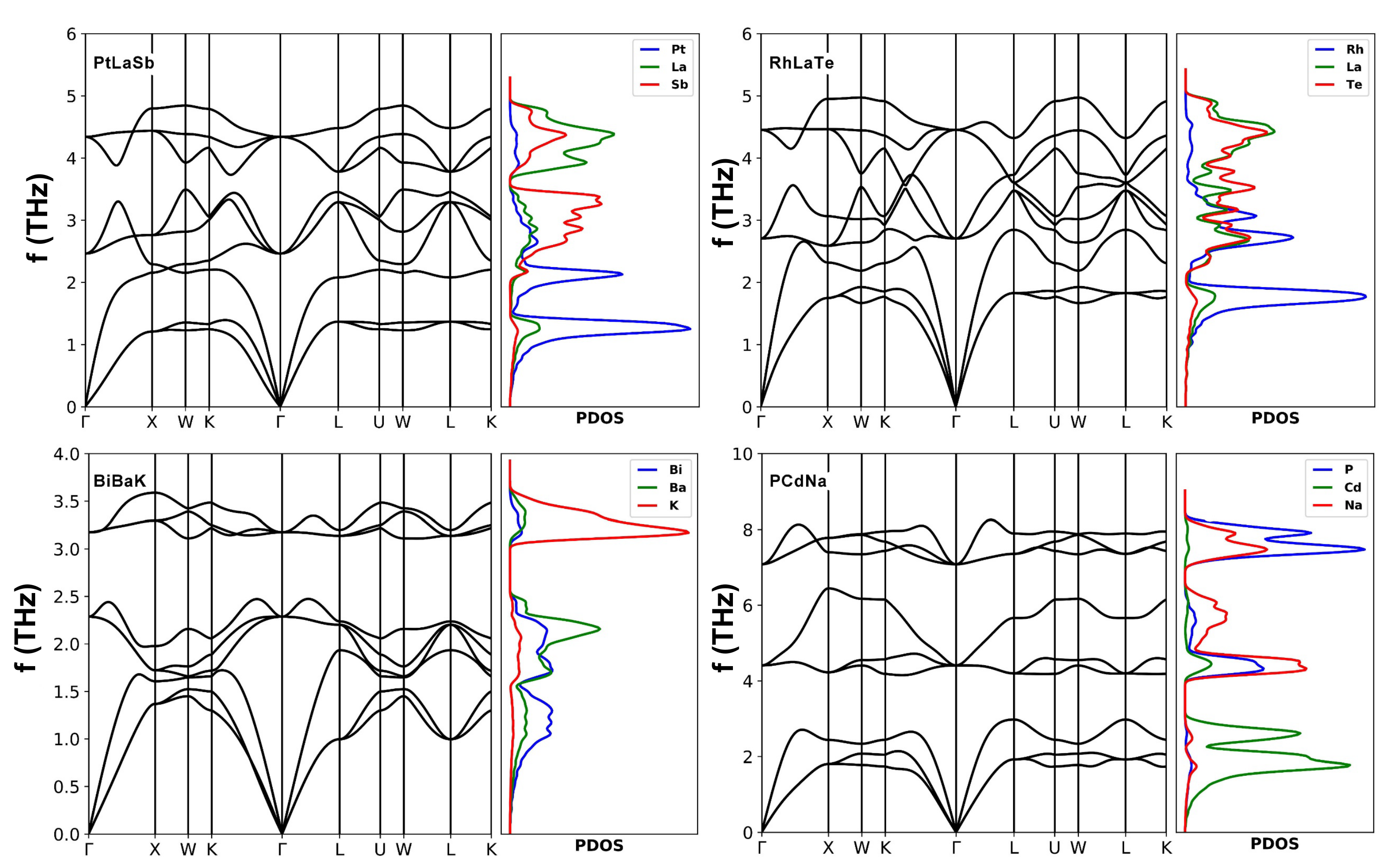}
\caption{\label{figure5}
Phonon dispersions and projected phonon density of states for
PtLaSb, RhLaTe, BiBaK and PCdNa.
Note the different frequency scales for different compounds.}
\end{figure*}
 
The ratios of $k_{min}$ to $k_{max}$ are shown in Fig. \ref{figure2}(b).
Here $k_{min}$ and $k_{max}$ are the smallest and largest
effective spring constants among the three atoms.
It is useful to remember that the half-Heusler structure can be regarded
as a filled zinc blende structure. In the two atom zinc blende structure
there is an expectation that the effective spring constants for the two
atoms are generally similar. This is because they would necessarily be
identical only nearest neighbor interactions were present.
Small ratios of $k_{min}$ to $k_{max}$ mean that one atom is weakly
bonded relative to the the others. 
Weak bonding of an atom will lower the sound velocity because the average
stiffness of the lattice will be reduced, which can be expected to reduce
the thermal conductivity.
It will also lead to an atom whose motion becomes decoupled from the other
atoms, leading to low frequency Einstein-like phonon branches.
This connects intuitively with the concept of rattling, which as mentioned
can have strong effects on the thermal conductivity.

Crystallographic atomic displacement parameters
(ADP, i.e. from experimental Debye-Waller factors), and calculated MSD
have been discussed as an indicator of low thermal conductivity.
\cite{sales-1999,nolas2000}
The ADP at a given
temperature is determined by the effective spring constant, and not
the frequency, since heavier mass will lower the frequency but will not
increase the ADP.

We now discuss some other quantities that are commonly discussed
in the context of thermal conductivity.
The average frequency and the ratio of
$\omega_{min}$ to $\omega_{max}$ are shown in Fig. \ref{figure2}(c) and (d).
Not surprisingly there is a correlation between the
average phonon frequency and $\kappa_l$.
Importantly, there is also a clear correlation of $\kappa_l$ with the ratio.
This relates to the
idea of rattling in terms of an atom giving low frequency optic modes.
Specifically, the ratio is not connected with the average
sound velocity and therefore would naively not be a key parameter from the
viewpoint of the Callaway expression.

The Debye temperature is another measure related to average phonon
frequency. There are different definitions of the Debye temperature
related to what quantity is being measured, e.g. specific heat, mean
square displacements (x-ray Debye temperature) etc. Considering that
thermal conductivity is often discussed in terms of acoustic modes,
we calculated the elastic Debye temperature, which is the specific
heat Debye temperature given by the elastic constants. This Debye
temperature would not be affected by low lying optic modes such as
those introduced by rattlers, except to the extent that weakly bound
atom would reduce the overall elastic stiffness of the lattice. We
obtained the the Debye temperature $\Theta$, and the average sound
velocity $v_m$. The correlation of $\kappa_l$ with these quantities.
is shown in Fig. \ref{figure3}.
The Callaway expression applied to acoustic modes would suggest
a strong correlation between average velocity and $\kappa_l$.
Our results show that,
while correlated, this not as strong a correlation as one might expect.
In fact, it is similar to the correlation with $\Theta$, although
the $\Theta$ also has a dependence on lattice parameter.

\begin{table*}[tbp]
\caption{Phonon frequency $f$(THz),
the spring constant $k$ (N/m) for each atom,
the ratio between the minimum to maximum phonon frequency
and spring constant in PtLaSb, RhLaTe, BiBaK, PCdNa.}
\label{4-compound}\tabcolsep0.01in
\begin{ruledtabular}
\renewcommand{\arraystretch}{1.2}
\begin{tabular}{cccccccccccccc}
  
Comp.   & $f_{1}$  &$f_{2}$ & $f_{3}$ &$f_{average}$ & $f_{min} / f_{max}$ & $k_{1}$  &$k_{2}$ & $k_{3}$ &$k_{average}$ &$k_{min}/k_{max}$ & $\kappa_{l}$\\ \hline
   PtLaSb   &1.81 &3.60 &3.27 &2.91 &0.50 &41.83 &117.98 &85.33  &81.71  &0.35 &0.84    \\ 
   RhLaTe   &2.32 &3.47 &3.52 &3.12 &0.66 &36.22 &109.79 &103.74 &83.25  &0.33 &1.21     \\ 
   BiBaK    &1.62 &1.96 &3.02 &2.21 &0.54 &36.06 &34.54  &23.41  &31.34  &0.65 &2.26      \\   
   PCdNa    &6.53 &2.31 &5.59 &4.85 &0.35 &86.60 &39.41  &47.04  &57.68  &0.46 &2.27       \\ 
\end{tabular}
\end{ruledtabular}
\end{table*}

\begin{figure}
\includegraphics[width=0.8\columnwidth]{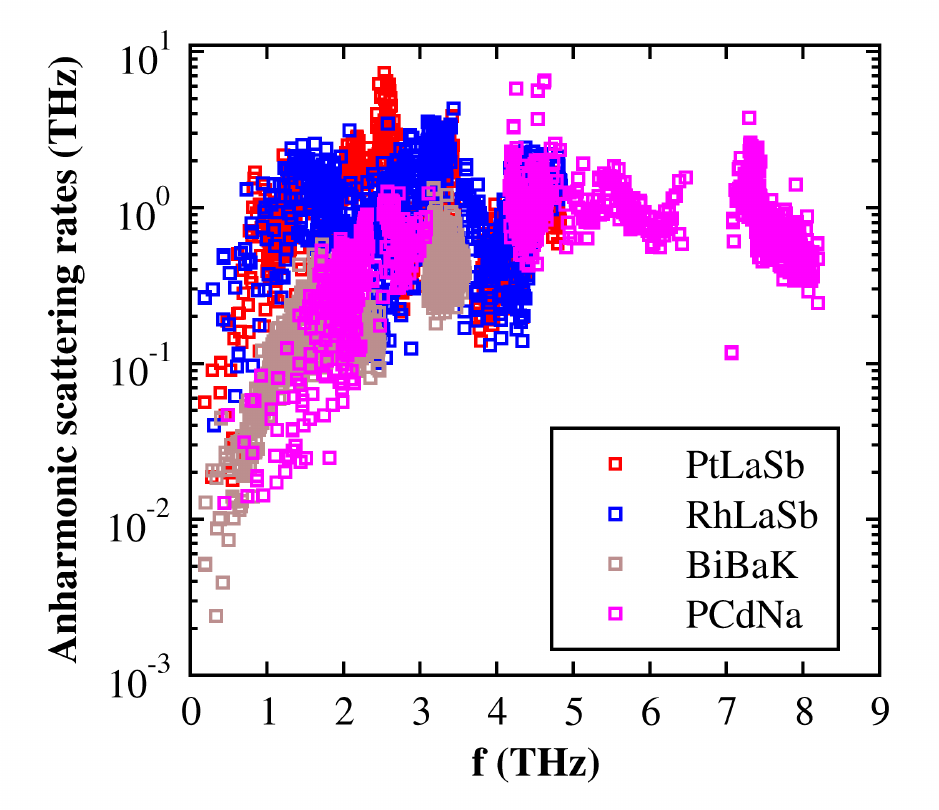}
\caption{\label{anharmonic}
Calculated 300 K anharmonic scattering rates for the four low thermal
conductivity compounds.}
\end{figure}

From our calculations,
we find that PtLaSb, RhLaTe, BiBaK, and PCdNa have very low
lattice thermal conductivity.
The phonon dispersions and the projected phonon densities of states of these
four compounds are given in Fig. \ref{figure5}.
The average phonon frequencies
for each atom and the effective
spring constants $k_\alpha$ for these
four compounds are listed in Table \ref{4-compound}.
The phonon dispersions of these four compounds show a strong difference
between transverse and and longitudinal acoustic modes. In other
words the transverse acoustic branches have much lower velocity than the
longitudinal branches.
While this is a commonly observed characteristic of materials, it is
not a general characteristic of the zinc blende structure, from which
the half-Heusler structure is derived. In the zinc blende semiconductors,
the transverse branches are typically stiff reflecting bond bending
forces from covalent bonding. This is known to lead to higher thermal
conductivity, as in BaAs,
\cite{lindsay2013}
while conversely materials that have low
velocity transverse branches, such as CuCl, generally have low thermal
conductivity.
\cite{mukhopadhyay}

The phonon densities of states for the four compounds indicate that the
low-energy acoustic modes are dominated by one atom (Pt/Rh/Bi/Cd) vibrations,
whereas optical branches are governed by another two atoms.
Moreover, there are avoided crossings
of the longitudinal acoustic branch and optical phonon branches, which
are a characteristic of rattling.
The anharmonic scattering rates, which include
both the effect of anharmonicity and the scattering phase space
are clearly important for thermal
conductivity, These are shown in Fig. \ref{anharmonic} for the
four compounds.
The scattering rates for low frequency phonons are highest for the
lowest thermal conductivity materials, specifically PtLaSb and RhLaSb,
as may be expected.
Finally,
the ratios between the smallest to largest effective spring constant for
PtLaSb, RhLaTe, BiBaK, and PCdNa are low,
0.35, 0.33, 0.65, and 0.46, respectively.

\section{Summary and Conclusions}

We investigated the thermal conductivity of half-Heusler semiconductors
in relation to the phonon dispersions using different measures related
in particular to rattling. We find that the thermal conductivity is 
correlated with average phonon frequency as expected and also surprisingly
well with average effective spring constant. This is connected with the
idea that weak bonding leads to greater anharmonic scattering.
We constructed two measures based on local dynamics using the
site average phonon frequency from the projected phonon density of states.
The first is a ratio of the lowest site average frequency to the highest.
The second is a ratio of the lowest effective spring constant to the highest.
Both of these correlate with thermal conductivity and are different from
each other. For identifying the lowest thermal conductivity the first
($\omega_{min}/\omega_{max}$) is somewhat better in this set of compounds.
This measure corresponds to the idea that low rattling frequency is best.
The other measure ($k_{min}/k_{max}$), which measures bonding is also
well correlated with thermal conductivity. We note that neither of these
ratios scales with phonon velocity.
We hope that these results are useful in providing understanding of rattling
in relation to thermal conductivity and perhaps in screening materials for
potential low thermal conductivity.

\acknowledgments

Work at the University of Missouri was supported by
the U. S. Department of Energy, Office of Science, Basic Energy Sciences,
Award No. DE-SC0019114. 
Work at ISSP and USTC was supported by National Science Foundation of
China, Award 11774347. 
Z.F. gratefully acknowledges support from the China Scholarship Council (CSC).

\bibliography{kl.bib}

\end{document}